# TO ACHIEVE MAXIMAL THROUGHPUTS IN CSMA WIRELESS NETWORKS THROUGH OFFERED-LOAD CONTROL

Caihong Kai, Qi Zhang and Lusheng Wang

School of Computer and Information, Hefei University of Technology, Hefei, China

Email: chkai@hfut.edu.cn, islandqi@126.com

**Abstract**—This paper studies how to achieve the maximal link throughputs in a CSMA wireless network through offered-load control. First, we propose an analytical model, contention-graph-combination (CGC), to describe the relationship between the offered-load and the output link throughputs of an unsaturated CSMA network. Based on CGC, we then formulate a linear optimization model to improve the aggregate link throughput through properly setting the occurrence probabilities of each sub-network, based on which we can obtain the optimal offered-load of each link. Simulation results bore out the accuracy of our CGC analysis and the maximal link throughputs can be closely achieved. Different from prior work in which CSMA protocol parameters are adaptively adjusted to achieve better performance, in this paper we propose to achieve maximal link throughputs by adjusting the rates of the traffic pumped into the source nodes of links, which runs in a software manner and is more practical to implement in real networks.

## I INTRODUCTION

This paper concerns the throughput maximization problem of a carrier-sense multiple access (CSMA) wireless network in which links compete with each other for airtime usage through the CSMA protocol. When a link senses a packet transmission on the channel, it will not attempt to transmit and perform random backoff process in order to prevent packet collisions.

Due to the intricate and inseparable interactions and dependencies among links in the network, the throughput analysis and optimization problem is difficult to solve and has attracted much attention from the research community [1]-[9]. Analytical models were developed to characterize the behaviors of links in channel competitions and compute the throughput each link obtains [1]-[4]. Moreover, optimization algorithms were proposed to achieve better system performance for CSMA networks [5]-[8].

Different from most of prior work in which the saturated networks are considered, in this paper we study the throughput maximization problem given a general offered-load at the sources of links in the network. More specifically, we first build up an analytical model, contention-graph-combination (CGC), to characterize the relationship between the offered-load and the link throughput distributions in the network. The basic idea of CGC is as follows: when a link has no packet to transmit, it will quit channel competition until the next packet arrival. From the perspective of the network topology, we can regard that the link is turned "off" and the link will be turned "on" again upon the next packet arrival on its transmit buffer . Hence, the network topology varies according to the "on" and "off" of the unsaturated links. Thus, a CSMA network can be regarded as operating over a set of different contention graphs with saturated traffics and can be analyzed based on saturated results.

More importantly, based on CGC we further formulate an optimization problem to study how to achieve the maximal aggregate link throughputs through appropriate offered-load settings. That is, different from previous work in which the CSMA protocol parameters are adaptively adjusted to achieve better performance, we optimize network performance by properly setting the rates of traffics that are pumped into the transmit buffer of the sources. Simulations validate the accuracy of our CGC model and show that the aggregate link throughputs can be largely increased through proper offered-load control.

It is worthwhile to note that from the viewpoint of engineering implementation, our method is superior to previous algorithms because it is much easier to control the input rates of links at the MAC layer than to adjust the CSMA parameters the links adopt for channel competition. Recall that the standard CSMA parameters have been fixed in the firmware of typical commercial wireless cards (e.g., IEEE 802.11a/b/g wireless cards[9]), to adaptively adjust these parameters are not easy to implement in practical wireless networks.

**Related Work**

With the increasingly deployment of CSMA wireless networks (e.g., Multi-cell WLAN, wireless mesh and mobile Ad Hoc networks), researchers have been paying much attention to the analysis and optimization of CSMS networks [1]-[8]. In particular, the authors of [1] computed the link throughputs by a continuous-time Markov chain, based on which [2] studied the fairness problem in large CSMA networks. Recent work [3] proposed an idealized CSMA network (ICN) and theoretically proved that the Markov-chain computation is applicable for a general CSMA network even though the system process actually has memory. The authors of [4] proposed a uniform framework to capture the interactions among links and perform capacity analysis for CSMA networks. Furthermore, based on the ICN model, the authors of [5][6] proposed elegant adaptive algorithms (ACSMA) for CSMA wireless networks to achieve system utility maximization. Ref. [7] studied the applications of belief propagation in CSMA wireless networks and [8] proposed queue-length based CSMA/CA algorithms to achieve maximum throughput and low delay.

However, all works above assumed that each link has infinite traffic and the network is saturated[1]. Note that the traffic load in practical CSMA networks (e.g., WiFi and mesh networks) is finite and the transmit buffer of links can be empty from time to time [10]. As far as we know, how to optimize the CSMA network through offered-load control is outstanding in the literature and this paper attempts to bridge this gap.

It is known that due to the inseparable interactions and dependencies between links in the network, the behavior of an unsaturated link could affect the throughput each link obtains in the overall network [10]. Thus, it is non-trivial to build up the analytical connection between the output throughputs and the offered-load of links. This is the first thrust of this paper. Based on the established connection, we then can develop optimization model and analyze how to properly set the offered-load of links to achieve maximal aggregate throughputs.

The remainder of the paper is organized as follows. Section II introduces our system model and reviews the throughput analysis of saturated CSMA wireless networks. Section III presents our CGC model and the optimization problem of achieving maximal aggregate throughputs through offered-load control. Section IV shows extensive simulation results to examine the accuracy of our model and validate our optimization analysis. Finally, section V concludes the paper.

## II. SYSTEM MODEL AND SATURATED ANALYSIS

In this section, we introduce our system model of a CSMA wireless network with finite offered-load and briefly review the saturated analysis of ICN in [3].

### A. ICN Model with Finite Offered-load

We use a contention graph $G=(V,E)$ to describe the carrier sensing relationship among links in an ideal CSMA wireless network. Vertices are used to represent the links (a transmitter-receiver pair), and the edges are used to describe the carrier-sensing relationship among links (i.e., an edge joints two links if the associated transmitters can sense each other). We say two links are neighbors if there is an edge between them. Due to carrier sensing, the neighbors will prevent from transmitting at the same time to avoid packet collisions.

It senses the channel as idle if and only if none of its neighbors are transmitting. Under CSMA protocol, a link will not perform channel competition until it has a packet to transmit in its transmit buffer. Each link with a packet to transmit has a backoff timer, $t_{cd}$, whose initial value is a continuous variable with an arbitrary distribution. The timer decreases continuously with time when the channel is sensed idle. The backoff process is frozen once the channel becomes busy due to neighbors' transmission and the remaining backoff time is recorded. The backoff process resumes with the previous stored backoff time when the channel becomes idle. When the timer value decreases to zero, the link transmits a packet. Denote the packet transmission time by $t_{tr}$. After the transmission, the link will quit channel competition if its transmit buffer is empty. Otherwise, the timer is reset to a new random value and the above process repeats.

---

[1] In ACSMA proposed in [5], dummy packets are transmitted if a link has no packet waiting for transmission to make sure the network is saturated.

We let the access intensity of a link be the ratio between the mean packet transmission time and the mean backoff countdown time: $\rho = E[t_{tr}]/E[t_{cd}]$.

We assume that each link has an unlimited transfer buffer, and the packet arrival process satisfies Poisson distribution. Let $\overline{f} = [f_1, f_2, \cdots f_N]$ represents the offered-load of the overall network. We can use 0 and 1 to indicate the states of links, $s_i = 1$ denotes that link $i$ has a packet being transmitted and $s_i = 0$ denotes that link $i$ is in the countdown process or frozen. The system state of ICN is $s = s_1 s_2 \cdots s_N$. If link $i$ and link $j$ are neighbors, $s_i$ and $s_j$ can not both be 1 at the same time because they can sense each other, and the backoff time is a continuous random variable, thus the probability of that two links countdown to zero at the same time is zero.

### B. Saturated Throughput Computation under ICN

This part is a review of the saturated result in [3], and the reader is referred to [3] for more details. If we assume that the network is saturated and the backoff time and transmission time are exponentially distribution, then $s(t)$ is a time-reversible Markov process.

Let $S$ represent all possible states, and $\rho_i$ be the access intensity of link $i$. The stationary distribution of state $s = s_1 s_2 \cdots s_N$ is

$$P_s = \frac{\prod_{i:s_i=1 \text{ in } s} \rho_i}{Z}, \quad \forall s \in S \quad (1)$$

where $Z = \sum_{s \in S} \prod_{i:s_i=1 \text{ in } s} \rho_i$

We can get the normalized throughput of link $i$ as

$$th_i = \sum_{s:s_i=1} P_s \quad (2)$$

## III. CGC MODEL AND NETWORK THROUGHPUT MAXIMIZATION

In this section, we present our CGC model to analyze the relationship between the offered-load and the link throughputs. After that, we formulate our network throughput maximization problem in which the optimal offered-load is computed to achieve the maximal aggregate link throughputs.

### A. CGC Model for Unsaturated ICN

We consider the unsaturated CSMA network in which the source nodes of the links have finite traffics injected to their transmit buffer at the MAC layer and the transmit buffer could be empty from time to time if its output rate is higher than its input rate (i.e., the throughput it obtains in channel competition is higher than the offered-load at the sources).

Recall that in a CSMA network if a link has no packet to transmit, it will quit channel competition until the next packet arrival. Hence, the network contention graph varies according to the "joining" and "quitting" of channel competition of links. Thus, a CSMA network can be regarded as operating over a set of different contention graphs. This is the basic idea behind CGC.

We next use an illustrating example to demonstrate the main idea of CGC. Consider the network shown in the left of Fig. 1, it is a four-link ring network. In the figure, a vertex represents a link and an edge joints two links if the transmitters of the associated links sense each other. That is, links 1 and 2 can hear links 3 and 4, links 3 and 4 can hear links 1 and 2, while links 1 and 2, links 3 and 4 cannot hear each other, respectively.

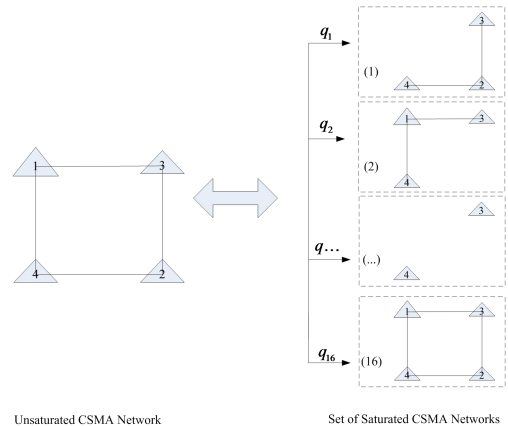

Unsaturated CSMA Network      Set of Saturated CSMA Networks

Fig. 1 An illustrating example to show the idea behind CGC

The left of Fig.1 shows the complete contention graph in which each packet joins channel competition. However, under a general offered-load each link could be unsaturated. Let the input process of each link be an "on-off" process. During the "on" period, the input rate is 1 (normalized to the maximum throughput a link can obtain); during the "off" period, the input rate is 0 (no packet arrival during this period). Suppose that the "on" periods and "off" periods are long enough, then we can decompose the situation into $2^4 = 16$ equilibrium saturated scenarios. In this case, the link throughputs can be computed by combining its saturated throughputs under each sub-network.

In general, for an $N$-link network with finite offered-load, we can decompose it into $2^N$ saturated sub-networks according to the "joining" and "quitting" of channel competition of each link. Let $q_j$ denote the appearance probability of sub-network $j$. The appearance probabilities of each such saturated sub-network is denoted by $\vec{q} = [q_1, q_2, \cdots, q_{2^N}]$.

Based on prior saturated analysis, we can build up the mathematical connection between the offered-load and link throughputs. The stationary distribution of state $s$ in sub-network $j$ is

$$P_s^j = \frac{\prod_{i:s_i=1 \text{ in } s} \rho_i}{Z^j}, \quad \forall s \in S^j \quad (3)$$

where $S^j$ is all possible states of sub-network $j$ and $Z^j = \sum_{s \in S^j} \prod_{i:s_i=1 \text{ in } s} \rho_i$. Then the throughput of link $i$ in sub-network $j$ can be computed by

$$th_i^j = \sum_{s:s_i=1} P_s^j \quad (4)$$

Thus far, we can compute the throughputs of link $i$:

$$th_i = \sum_{j=1}^{2^N} th_i^j q_j \quad (5)$$

As a proper offered-load, we set the offered load of each link to its achieved throughput. That is, $f_i = th_i, \quad \forall i \in V$.

B.  *Throughput Maximization Through Offered-load Control*

Based on the CGC model built up in Part A, we now formulate the optimization problem of throughput maximization through offered-load control.

Consider a CSMA wireless network in which each link has a lowest desired throughput $r_i$. Write $\vec{r} = [r_1, r_2, \cdots, r_N]$. That is, each link must guarantee a minimum throughput, which is determined by the application running on top of the wireless networks[2]. Through adjusting the offered-load $\vec{f}$, we attempt to achieve the maximal link throughputs given that the minimum throughput of each link is guaranteed. The optimization problem is formulated as

$$\text{objective function}: \quad \max_{\vec{q}} = \sum_{j=1}^{2^N} \left( \sum_{i=1}^{N} th_i^j \right) q_j \quad (6)$$

$$s.t. \begin{cases} \sum_{j=1}^{2^N} th_i^j q_j \geq r_i, \forall i = 1, 2, \cdots, N & (7) \\ \sum_{j=1}^{2^N} q_j = 1, \ 0 \leq q_j \leq 1 & (8) \end{cases}$$

where objective function (6) maximizes the aggregate link throughputs; constraint (7) describes that each link has a minimum required throughput and (8) is that the summation of the possibilities of all sub-networks is 1.

Recall that the link throughputs of each sub-network can be computed by (4), the above optimization problem is a linear optimization problem, which can be solved by standard methods under the optimization framework. Alternatively, in our problem, we have $2^N$ variables to be determined while have $N+1$ constraints. Thus, it is known that the optimal $\vec{q}$ has at most $N+1$ non-zero elements. Making using of this property, we can design heuristic algorithms to compute the optimal $\vec{q}$. The details will be presented somewhere else to limit the scope of this paper.

Thus far we have worked out an optimal $\vec{q}$ to make the network achieve the maximum aggregate throughputs. We denote the maximal link throughput under optimal $\vec{q}$ by $\vec{th}^* = [th_1^*, th_2^*, \cdots, th_N^*]$, which can be computed by (5). The optimal offered-load $\vec{f}^*$ is then set to $\vec{th}^*$, under which the maximal aggregate link throughputs can be achieved.

In our example shown in the left of Fig.1, let the required throughput be $\vec{r} = [0.1994, 0.3779, 0.4263, 0.4271]$. By solving the linear optimization problem, the maximum throughput of networks is $\overline{th}^* =[0.4261, 0.4261, 0.4271, 0.4271]$ and we set it to be the offered-load $\overline{f}^*$. We then set up simulations to simulate the CSMA protocol by MATLAB programs. Given the offered-load $\overline{f}^*$, we find that the output throughputs of the network is very close to [0.4261, 0.4261, 0.4261, 0.4261], indicating that our CGC model for unsaturated ICN analysis is of high accuracy and the maximal aggregate link throughputs can be achieved by optimal offered-load control. More simulation results will be presented in Section IV.

---

[2] To guarantee that the desired minimum throughput is achievable, we assume that the given $\vec{r}$ is strictly feasible according to the definition in [5]. For example, $r_i \leq \dfrac{\rho_i}{1+\rho_i}$, which is the maximum throughput of link $i$.

## IV. SIMULATION EVALUATION

We conduct simulations to examine the accuracy of our CGC analysis and the offered-load optimization computation. We implement an ICN-simulator with finite offered-load using MATLAB programs. The optimal offered-load of our linear optimization model is computed by the standard *linprog* function in MATLAB. Typical link access intensity $\rho_0 = 5.3548$ is used in the simulations.

We solve for the optimization problem to get the maximal aggregate link throughputs $\overline{th}^*$ and the optimal offered-load $\overline{f}^*$, and then examine the output link throughputs $\overline{th}$ of a CSMA network under the optimal offered-load.

In Table I we compare the link throughput calculated by the optimization model with the throughput obtained from the ICN-simulator for different topologies and offered-load vectors. As can be seen there, the link throughputs obtained from the ICN-simulator are very close to the link throughputs computed by the *linprog* function. This indicates: 1) Our analysis on unsaturated CSMA network, CGC model, has high accuracy; 2) the maximal aggregate link throughputs can be perfectly achieved by setting the optimal offered-load.

TABLE I Contention Graphs of Various Network Topologies, Minimum Link Throughputs, Optimal Offered-load and the Simulated Link Throughputs under Optimal Offered-load

| Network | $\overline{r}$ /sum($\overline{r}$) | $\overline{f}^*$/sum($\overline{f}^*$) | $\overline{th}$ /sum($\overline{th}$) |
|---|---|---|---|
| 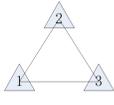 | [0.0998, 0.3510, 0.1999] | [0.2854, 0.3510, 0.2980] | [0.2850, 0.3506, 0.2971] |
| | 0.6507 | 0.9344 | 0.9327 |
| 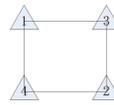 | [0.1994, 0.3779, 0.4263, 0.4271] | [0.4261, 0.4261, 0.4271, 0.4271] | [0.4261, 0.4260, 0.4264, 0.4265] |
| | 1.4307 | 1.7064 | 1.705 |
| 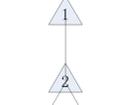 | [0.4997, 0.0997, 0.4118, 0.4116] | [0.6828, 0.0997, 0.4118, 0.4116] | [0.6827, 0.0990, 0.4114, 0.4118] |
| | 1.4228 | 1.6059 | 1.6049 |
| 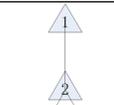 | [0.5004, 0.0204, 0.8250, 0.8250] | [0.8254, 0.0204, 0.8254, 0.8254] | [0.8254, 0.0205, 0.8243, 0.8252] |
| | 2.1708 | 2.4966 | 2.4954 |
| 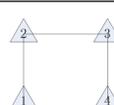 | [0.5998, 0.1999, 0.2004, 0.5778] | [0.5998, 0.2882, 0.3143, 0.5778] | [0.5996, 0.2881, 0.3142, 0.5775] |
| | 1.5779 | 1.7801 | 1.7794 |

For larger networks, we run three sets of 10 randomly generated 10-link networks. **Setting 1:** we generate networks with different mean degree of links (i.e., the number of neighbors per link in the contention graph) to around 2, 3 and 4. **Setting 2:** for the 10 networks with mean link degree to 2, we adjust the access intensities of links. **Setting 3:** for the 10 networks with mean link degree to 2, we adjust the minimum required link throughputs.

To make sure that minimum required throughput $\overline{r}$ is feasible (otherwise, the optimization problem has no solution), we first compute the link throughputs under the saturated scenario (i.e., each link has infinite source traffic) and then set the minimum required throughput to be no more than the saturated throughput. More specifically, in Settings 1 and 2, for odd links, we set $r_i = \max(th_i^0 - 0.2, 0)$; for even links, we $r_i = \max(th_i^0 - 0.1, 0)$, where $th_i^0$ is the saturated throughput of link $i$.

For each network, we collect the gap between the maximal link throughput $\overline{th}^*$ and the achieved link throughput $\overline{th}$. Define the mean link throughput errors to be $mean(|\overline{th}^* - \overline{th}|)$ and the aggregate link throughput gap to be $\left|\sum_{i=1}^{N} th_i^* - \sum_{i=1}^{N} th_i\right|$.

Table II, III and IV show the mean link throughput errors and the mean aggregate link throughput gap for the three simulation settings above, respectively. In the table, each value is averaged over 10 random networks. As can be seen there, the simulated link throughput is very close to the computed maximal link throughput under all the scenarios we test. The mean link throughput error as well as the mean aggregate throughput error is kept below 1% for all test scenarios. Thus, our offered-load control can make sure the CSMA network perfectly achieve the maximal throughputs.

TABLE II Mean Link/Aggregate Throughput Errors for networks with different mean link degrees

| Mean Link Degree | 2 | 3 | 4 |
|---|---|---|---|
| Mean Link Throughput Errors | 0.20% | 0.33% | 0.37% |
| Mean Aggregate Throughput Errors | 0.39% | 0.83% | 1.09% |

TABLE III Mean Link/Aggregate Throughput Errors for networks with different access intensities

| Access Intensity | $\rho_0$ | $2\rho_0$ | $3\rho_0$ |
|---|---|---|---|
| Mean Link Throughput Errors | 0.20% | 0.15% | 0.11% |
| Mean Aggregate Throughput Errors | 0.39% | 0.27% | 0.19% |

TABLE IV Mean Link/Aggregate Throughput Errors for networks with different minimum required link throughputs

| | $\bar{r}$ | max($r_i$ -0.1, 0) | max($r_i$ -0.1, 0) |
|---|---|---|---|
| Mean Link Throughput Errors | 0.20% | 0.10% | 0.069% |
| Mean Aggregate Throughput Errors | 0.39% | 0.19% | 0.13% |

### V. CONCLUSION

This paper proposed an analytical model, CGC, to analyze the link throughputs of a CSMA network with finite offered-load. Based on CGC, we built up an optimization model to achieve maximal aggregate link throughputs through offered-load control. Simulation results showed that our analysis is of high accuracy and the maximal link throughputs can be achieved under various network and parameter settings.

One remaining issue of our CGC analysis and the optimization model is that due to the exponential increase of the number of the sub-networks with respect to the number of links, the optimization problem is of high complexity of solve. In the future work, we will further investigate how to design quick and accurate algorithms to solve the optimization problem and make them suitable for practical CSMA wireless networks.